\documentclass{ws-mplb}
\usepackage{amsfonts,amssymb,amsmath}
\usepackage{graphicx}
\usepackage[super,sort,compress]{cite} 
\usepackage{shortcuts}

\begin{document}

\markboth{\DJ or\dj e Radi\v cevi\'c}{Entanglement entropy across the lattice-continuum correspondence}

%
%

\title{Entanglement entropy across the lattice-continuum correspondence}

\author{\footnotesize \DJ or\dj e Radi\v cevi\'c}

\address{Martin Fisher School of Physics, Brandeis University\\
Waltham, MA 02453, USA
\\
djordje@brandeis.edu}

\maketitle

\vspace*{2em}

\begin{abstract}
This paper revisits standard calculations of free field entanglement entropy in light of the newly developed lattice-continuum correspondence. This correspondence prescribes an explicit method to extract an approximately continuum quantum field theory out of a fully regularized lattice theory. This prescription will here be extended to subregion algebras, and it will be shown how entropies of continuum boson and fermion theories can be computed by working purely with lattice quantities. This gives a clear picture of the origin of divergences in entanglement entropy while also presenting a concise and detailed recipe for calculating this important quantity in continuum theories.
\end{abstract}

\keywords{Quantum field theory; entanglement entropy; continuum limit.}

\section{Introduction}

There are many ways to quantify the entanglement between different parts of a quantum system in a given state. One probe of great interest is the entanglement entropy (EE) associated to a part of the original system\cite{Bombelli:1986rw, Srednicki:1993im, Callan:1994py, Vidal:2002rm, Somma:2004}. This quantity measures how much information about the starting state becomes lost if our knowledge is restricted to correlation functions only within the chosen subsystem\cite{Zanardi:2004zz}. The dependence of the ground state EE on the subsystem size and shape typically encodes many salient properties of the quantum theory in question.

If the system under study is a quantum field theory (QFT) on a $d$-dimensional space $\Mbb$, the traditional object of interest is the EE associated to a spatial subregion $\Vbb \subset \Mbb$. It is now widely appreciated that there are many inequivalent ways to precisely define this quantity \cite{Casini:2013rba, Radicevic:2014kqa}. Each choice of an algebra $\A^\alpha_\Vbb$ of operators that are at least approximately supported on $\Vbb$ --- and each choice of a representation of this subalgebra --- generically leads to a different entropy $S^\alpha_\Vbb$. Most results in the vast existent literature on entanglement hold for the maximal operator algebra $\A\^{max}_\Vbb$ taken in a certain natural representation \cite{Lin:2018bud}.

This paper will focus on an even more fundamental choice needed to discuss QFT entanglement --- the precise definition of the QFT itself. Broadly speaking, there are three different ways to set up the QFT framework for the study of EE: \newpage
\begin{romanlist}
  \item \bt{The lattice approach.} Let $\Mbb$ be a finite lattice whose elements (sites, links, plaquettes\ldots) host quantum degrees of freedom with finite-dimensional Hilbert spaces. This defines the QFT as a finite (but possibly very large) quantum system. For each QFT density matrix $\rho$ and a choice of algebra $\A^\alpha_\Vbb$ supported on the sublattice $\Vbb \subset \Mbb$ there is now a unique reduced density matrix $\rho^\alpha_\Vbb \in \A^\alpha_\Vbb$. The desired EE is the von Neumann entropy of this density matrix,
      \bel{\label{def S trace}
        S_\Vbb^\alpha(\rho) = - \Tr \left[\rho^\alpha_\Vbb \log \rho^\alpha_\Vbb \right]
      }
      Note that the trace depends on the representation of the operators in $\A_\Vbb^\alpha$.
  \item \bt{The everyday continuum approach.} Let $\Mbb$ be a smooth manifold, and let $\Ebb \equiv \Mbb \times \R$ be the associated Euclidean spacetime. There are many ways to define an effective continuum QFT on it, e.g.\ by starting from a classical action on $\Ebb$ and quantizing it, or by placing a conformal QFT on $\Ebb$ and perturbing it by relevant operators. Either way, this results in some kind of self-consistent set of rules for calculating QFT  partition functions in the presence of various background fields. Then the ground state EE of the region $\Vbb \subset \Mbb$ can be defined as
      \bel{\label{def S replica}
        S^\alpha_\Vbb(\trm{vacuum}) = - \lim_{n \rar 1} \pder{}n \log \frac{\Zf^\alpha(\Ebb^n_\Vbb)}{\Zf(\Ebb)^n},
      }
      where $\Zf(\Ebb)$ is the regular partition function on $\Ebb$, and $\Zf^\alpha(\Ebb^n_\Vbb)$ is the partition function of the QFT on the manifold $\Ebb^n_\Vbb$ obtained from $\Ebb$ by replicating the region $\Vbb$ precisely $n$ times.\cite{Calabrese:2004eu} The most interesting algebra choices $\alpha$ correspond to different boundary conditions at the conical singularities along the edges of $\Vbb$, and choosing different representations corresponds to including certain decoupled sectors of fields in the partition function \cite{Lin:2018bud}. In the special case in which the QFT in question has a classical holographic dual, the EE associated to the maximal algebra $\A_\Vbb\^{max}$ is equal to the area of the Ryu-Takayanagi surface homologous to $\Vbb$ in the dual theory on AdS$_{d + 2}$ \cite{Ryu:2006bv}. \\
      \hspace*{1em} A characteristic of every calculation along these lines is that the EE contains cutoff-dependent (often divergent) terms with no physical meaning in the effective QFT. One must take care to isolate only the \emph{universal} part of the EE.
  \item \bt{The rigorous continuum approach.} Take $\Mbb$ to be a smooth manifold as before, but define the QFT using a rigorous operator-algebraic framework, e.g.\ following Haag and Kastler \cite{Haag:1964}. This ensures that all quantities that are well defined in the first place will turn out to be finite. Here it is again possible to define a multitude of  algebras $\A_\Vbb^\alpha$ associated to a spatial subregion $\Vbb$ (or to its causal diamond in spacetime). Instead of working with an ill defined quantity like a partition function, however, quantum information is encoded by the relative Tomita operator \cite{Witten:2018zxz}, determined by the relations
      \bel{
        \mathcal T^\alpha_\Vbb \O \qvec{\Psi} = \O\+ \qvec{\Phi}
      }
      for every operator $\O \in \A_\Vbb^\alpha$ and for two fixed states $\qvec \Psi$, $\qvec \Phi$. The well defined analog of EE is now the \emph{relative entropy} of these states, which is given by
      \bel{\label{def S Tomita}
        S^\alpha_\Vbb(\Psi|\Phi) = - \qmat\Psi {\log \left[(\mathcal T^\alpha_\Vbb)\+ \mathcal T_\Vbb^\alpha\right]} \Psi.
      }
\end{romanlist}

It may seem that the rigorous continuum approach in some sense cheats by replacing EE with relative entropy. This is actually not a very momentous change. The relative entropy between two states can also be defined within the lattice and everyday continuum approaches, and it can be used to diagnose QFT properties in much the same way as ordinary EE. Indeed, this paper could have just focused on the relative entropy from the beginning. This path was not taken because EE is already ubiquitous in the literature --- unlike the rigorous continuum approach.

Having come out in support of studying ordinary EE, this paper will henceforth disregard traditional algebraic QFT and focus solely on the two approaches that do compute the EE itself. But what is the precise relation between entropies calculated by formul\ae\ \eqref{def S trace} and \eqref{def S replica}?

It is well known that lattice QFTs with long-range correlations can at low energies be expressed as effective continuum theories \cite{Kadanoff:1970kz, Wilson:1971bg}. In fact, many examples demonstrate that universal continuum data, such as the central charge, can be extracted from lattice calculations of the EE using eq.\ \eqref{def S trace} \cite{Holzhey:1994we, Vidal:2002rm}. It is also known, however, that continuum QFTs do not have factorizable Hilbert spaces, and so the partial trace used to write \eqref{def S trace} is not meaningful in the continuum. Conversely, while it is possible to perform the replica trick on lattice systems, there is in general no rigorous connection between the resulting discrete path integrals and the continuum path integrals that figure in eq.\ \eqref{def S replica}. Until now, the best one could do was to show via an explicit calculation that judiciously chosen pairs of continuum and lattice path integrals have the same universal parts of the EE.

The goal of this paper is to provide a set of rigorous tools for understanding when and how the two approaches to EE actually agree. The idea is to use the recently developed methods that start from a lattice theory and ``smoothe it out'' to give an effective continuum theory \cite{Radicevic:2019jfe, Radicevic:2019mle, Radicevic:1D, Radicevic:2D, Radicevic:3D}. It is already known how many properties of continuum QFTs --- Lorentz-invariant actions, operator product expansions, current algebras, infrared dualities --- explicitly emerge from the lattice within the smoothing framework. Here this list will be extended by finding natural subregion algebras of smooth lattice fields whose entropies take the familiar continuum forms.

The proposal for the lattice realization of continuum subregion algebras given here is new and applicable to all continuum field theories. Nevertheless, the approach will be primarily pedagogical. The aim is to convince the reader, via several worked out examples, that this proposal indeed naturally gives rise to a computable connection between lattice and continuum EEs.

The examples studied here will be the free fermion and boson theories in $(1+1)$D. Though these models are quite simple, they hide a wealth of interesting phenomena. Here these continuum QFTs will be defined completely rigorously, and the steps leading to the requisite EEs will be quite explicit.
The approach taken here will be operator-algebraic from start to finish. Path integral techniques will not be used, but the end results will be compared to existing results that were obtained using path integrals.

\newpage

\section{A review of smoothing}

What follows is a brief but self-contained review of the smoothing framework \cite{Radicevic:2D}. In order to showcase the breadth of this approach, this Section is more abstract than the rest of this paper. Readers who prefer to keep things concrete are invited to pick their favorite lattice theory, e.g.\ a free noncompact scalar on a cubic lattice, and merely verify that it can be smoothed as described below.

Let the \emph{original space} $\overline\Mbb$ be a finite lattice whose elements host the QFT degrees of freedom, with the dynamics governed by a Hamiltonian $H$. Two conditions must be satisfied for this lattice theory to have a continuum description at low energies:
\begin{romanlist}
  \item \bt{A precontinuum basis.} There must exist a complete set of commuting operators $\{n_k\}_{k \in \Pbb}$ that commute with $H$ and that can be diagonalized to
      \bel{\label{def nk}
        n_k = \trm{diag}(0, 1, \ldots, J_k - 1) \otimes \bigotimes_{k' \in \Pbb \backslash \{k\}} \1_{k'} \quad \trm{for}\quad J_k \geq 2.
      }
      The labels $k$ will be called \emph{momenta}, and their collection $\Pbb$ --- the \emph{momentum space}. \emph{Ladder operators} $c_k\+$ and $c_k$ change the \emph{particle numbers} $n_k$ by $\pm 1$, and their normalization can be chosen so that $n_k = c_k\+ c_k$. The set of all possible products of ladder operators is a basis of the QFT operator algebra that will be referred to as the \emph{precontinuum basis}. This is a generalization of the familiar Fock basis. Importantly, here there is no notion of particle statistics, and $\Pbb$ need not be the Fourier dual of $\overline \Mbb$.
  \item \bt{An induced metric on $\Pbb$.} It must also be possible to introduce a distance between momenta (i.e.\ a lattice structure on momentum space) such that the energy cost of changing particle numbers at momentum $k$ varies slowly with $k$. For simplicity, a more restrictive condition will be assumed here --- that the Hamiltonian takes the form
      \bel{\label{def H canonical}
        H = h^{(0)} + \sum_{k \in \Pbb} h^{(1)}_k n_k + \trm{small perturbations},
      }
      and that a lattice structure on $\Pbb$ can be chosen such that the \emph{dispersions} $h^{(1)}_k$ are uniformly continuous w.r.t.\ the natural lattice metric $||\cdot||$, i.e.\ such that
      \bel{
        \big| h^{(1)}_k - h^{(1)}_l \big| \propto ||k - l||
      }
      for any $k, l \in \Pbb$ whose distance is sufficiently small compared to the size of $\Pbb$. In other words, the theory will be assumed to look approximately free in the precontinuum basis, with a dispersion that is a nice function of the momentum. For this to make sense the momentum space must be large, $|\Pbb| \gg 1$.
\end{romanlist}

One important property of this setup is that the notions of particles, dispersions, etc.\ are all derived from the starting Hamiltonian instead of being assumed \emph{a priori}. In fact, even the dimension $d$ of the lattice $\Pbb$ is determined by the dynamics and can be different from the dimension of the original space $\overline \Mbb$. The Fourier dual of the momentum space, a $d$-dimensional lattice denoted $\Mbb$, will be called \emph{the position space} and will not necessarily be equivalent to $\overline \Mbb$.

\newpage

The momentum space may be composed of several identical sublattices $\Pbb^i$. For example, this happens due to ``fermion doubling'' when the starting theory on $\overline \Mbb$ contains free fermions. In such scenarios it is very convenient to slightly alter the above notation: one of these sublattices will be labeled $\Pbb$, and the information on which sublattice a field belongs to will be recorded by an extra field index. Now one can view the theory as having multiple \emph{species} of quantum fields $c_k^i$ for $k \in \Pbb$.

The crucial idea can now be introduced. \emph{Smoothing} is the projection to the algebra of smooth operators $\A\^S$, whose basis is formed by all possible products of operators from the generating set
\bel{
  \{c_k^i, (c_k^i)\+\}_{k \in \Pbb\_S\^{tot}} \cup \{n_k^i\}_{k \notin \Pbb\_S\^{tot}}.
}
Here $\Pbb\^{tot}\_S \subset \Pbb$ is a subset of momenta associated to the smallest dispersions $|h^{(1)}_k|$. The operators from $\A\^S$ can thus only create or destroy particles of low energies. Since $h^{(1)}_k$ is a slowly varying function of $k$,  $\Pbb\_S\^{tot}$ can  be defined as a set of momenta in a small neighborhood of a ``Fermi surface'' $\Pbb_\star$,
\bel{
  \Pbb\_S\^{tot} \equiv \bigcup_{k_\star \in \Pbb_\star}\ \{k\}_{||k - k_\star|| \leq k\_S}.
}
The cutoff $k\_S \gg 1$ used here is much smaller than the linear size of the momentum space, which is a scale set by $N \equiv |\Pbb|^{1/d} \gg 1$. (When $\Pbb$ is a cubic lattice, $N$ is the number of sites in each direction.) In all examples in this paper, $\Pbb_\star$ will consist of a single point in $\Pbb$, and so $\Pbb\_S\^{tot}$ will simply be labeled $\Pbb\_S$ from now on.

The effect of smoothing is most striking when it is applied to position space fields $c_x^i$, $x \in \Mbb$.\footnote{
  Recall that $\Mbb$ is itself defined as the Fourier dual of $\Pbb$, meaning that its components $x$ label eigenfunctions $G^x_k$ of the graph Laplacian on $\Pbb$, and the Fourier transform is $c_x^i \equiv \sum_{k \in \Pbb} G_k^x c_k^i$. When $\Pbb$ is a cubic lattice of linear size $N$, the Fourier kernels can be chosen to be the familiar functions $G_k^x = N^{-d/2} \exp(\frac{2\pi\i}{N} k \cdot x)$.}
Their projections to the smooth algebra will be denoted $c^i(x)$. Assuming that $x$ and $x + e$ are neighboring sites, the smooth fields must obey
\bel{
  c^i(x + e) = c^i(x) + \del_e c^i(x) + O\left(k\_S^2/N^2 \right),
}
where $\del_e c^i(x)$ is a smooth operator whose entries are all $O(k\_S/N)$. Ensuring this kind of operator constraint is the main goal of the entire construction described so far. It is important to stress that the quantum fields $c^i(x)$ are still defined at each lattice site $x$, but they are no longer linearly independent.

A remarkable property of the smoothing construction is that operator product expansions (OPEs) emerge naturally. Given any two operators $\O_x$ and $\~\O_y$, their smoothing and multiplication do not commute with each other, and the ``commutator'' of these operations can be \emph{defined} to be their OPE,
\bel{\label{def OPE}
  \O_x \times \~\O_y \equiv \O\~\O(x, y) - \O(x) \~\O(y).
}
The OPE coefficients are thus set dynamically, by the expectations $\{\avg{n_k^i}\}_{k \notin \Pbb\_S}$ of the central generators of $\A\^S$. As a nontrivial example, the OPEs of the 2D Ising model can be computed this way without ever referring to conformal symmetry \cite{Radicevic:2019mle}.

\section{Free fermions}

The simplest example of smoothing is provided by the free Dirac fermion in $(1+1)$D \cite{Radicevic:2019jfe}. A convenient microscopic definition of this theory starts with a \emph{spinless} (i.e.\ one-component) fermion on a lattice $\overline \Mbb$ made up of $2N$ sites arranged in a circle. The Hamiltonian is
\bel{
  H
   =
  \i \sum_{x \in \overline\Mbb}
    \left( \psi_{x + 1}\+ \psi_x - \psi_x\+ \psi_{x + 1} \right),
}
where the fermions satisfy the usual commutation relations,
\bel{
  \{\psi_x, \psi_y\} = 0, \quad \{\psi_x, \psi_y\+\} = \delta_{xy} \1.
}
The precontinuum basis is obtained by Fourier-transforming the fermions via
\bel{
  \psi_x
   \equiv
  \frac1{\sqrt{2N}} \sum_{k = -N}^{N - 1} \psi_k\,\e^{\frac{2\pi\i}{2N} kx }.
}
In terms of the $\psi_k$'s, the Hamiltonian is
\bel{
  H = \sum_{k = -N}^{N - 1} \psi_k\+ \psi_k \, 2\sin\frac{\pi k}{N}.
}
As discussed in the previous Section, it is convenient to split $\{-N, \ldots, N - 1\}$ into two segments and to take the momentum space to be
\bel{
  \Pbb \equiv \left\{ -\frac N2, \ldots, \frac N2 - 1 \right\},
}
This gives two species of fermions that can be assembled into a \emph{Dirac spinor},
\bel{
  \Psi_k^\pm \equiv \bcol{\psi_k}{\psi_{k + N}}, \quad k \in \Pbb.
}
By introducing the particle number operators $n_k^\alpha \equiv (\Psi^\alpha_k)\+ \Psi^\alpha_k$ the Hamiltonian takes the form \eqref{def H canonical},
\bel{
  H = \sum_{k \in \Pbb} \left(n_k^+ - n_k^-\right) h_k, \quad h_k \equiv 2\sin\frac{\pi k}N.
}
It is now evident that the smoothing conditions are met, with $\Pbb$ being a circular lattice with $N$ sites and antiperiodic boundary conditions (because $h_{k + N} = -h_k$).

Smoothing the position space spinors
\bel{
  \Psi_x \equiv \frac1{\sqrt N} \sum_{k \in \Pbb} \Psi_k \, \e^{\frac{2\pi\i}N kx}
}
gives the continuum fields
\bel{
  \Psi(x) = \frac1{\sqrt N} \sum_{k \in \Pbb\_S} \Psi_k \, \e^{\frac{2\pi\i}N kx},
   \quad
  \Pbb\_S \equiv \{-k\_S, \ldots, k\_S - 1\}
}
for $1 \ll k\_S \ll N$. As anticipated by Haldane \cite{Haldane:1981zza}, the existence of the second cutoff $k\_S$ is key to a rigorous definition of many continuum hallmarks, such as fermion OPEs, Kac-Moody current algebras, and Abelian bosonization \cite{Radicevic:2019jfe}.
\newpage

This theory has four ground states labeled by the zero mode expectations $\avg{n^\pm_0}$. For concreteness, this paper will focus on the ground state characterized by
\bel{\label{Fermi sea}
  \avg{n_k^\pm} = \theta(\mp k), \quad k \in \Pbb,
}
where $\theta(k) = 1$ if $k > 0$ and $\theta(k) = 0$ if $k \leq 0$. This means that the $k = 0$ mode is unoccupied for either component (``chirality'') of the spinor $\Psi_k$. This choice does not significantly affect the EE, as will become apparent below.

In this ground state the two-point function of lattice fields is
\bel{\label{2 point}
  \avg{(\Psi_x^\alpha)\+ \Psi_y^\beta}
   \equiv
  \delta_{\alpha\beta} C^\alpha_{x, y}
}
where
\begin{align}
  C^+_{x, y}
   &=
  \frac1N \sum_{k = -\frac N2}^{-1} \e^{\frac{2\pi\i}N k(y - x)} \approx \frac1{2\pi\i} \frac{1 - (-1)^{x - y}}{y - x},\\
  C^-_{x, y}
   &=
  \frac1N \sum_{k = 1}^{\frac N2 - 1} \e^{\frac{2\pi\i}N k(y - x)} \approx \frac1{2\pi\i} \frac{1 - (-1)^{x - y}}{x - y},
\end{align}
with corrections of size $O(1/N)$ and $O(|x - y|/N)$. This result can be recorded as
\bel{\label{def C}
  C^\pm_{x, y} \approx \mp C_{x - y},
   \quad
  C_r \equiv \frac1{2\pi\i} \frac{1 - (-1)^r}r.
}
This ``singularity'' at $r \rar 0$ is also captured by the OPE $(\Psi^\alpha_x)\+ \times \Psi^\beta_y$ computed according to \eqref{def OPE}, as this OPE is governed by the $\avg{n_k^\alpha}$'s at high momenta \cite{Radicevic:2019jfe}.

Now pick an interval $\Vbb \subset \Mbb$ of $M$ consecutive sites. In order to set the notation (and expectations), it is useful to first review the standard lattice calculation of its EE. The maximal algebra $\A\^{max}_\Vbb$ associated to this region is generated by the set
\bel{
  \{\Psi^\alpha_x, (\Psi^\alpha_x)\+\}_{x \in \Vbb, \, \alpha \in \{\pm\}}.
}
(Note that this is not an anticommutator!) As described in the Introduction, the corresponding EE is defined as the von Neumann entropy of the reduced density matrix $\rho\^{max}_\Vbb \in \A\^{max}_\Vbb$ that reproduces all vacuum expectation values in $\Vbb$ via
\bel{\label{def rho V}
  \Tr \left[\rho\^{max}_\Vbb \O\right] = \avg \O, \quad \forall \O \in \A\^{max}_\Vbb.
}

Since the two chiralities are decoupled, the calculation is simplified by focusing on just the \emph{chiral algebra} generated by the set $\{\Psi^+_x, (\Psi^+_x)\+\}_{x \in \Vbb}$. (This algebra is perfectly well defined; there is \emph{absolutely nothing} inconsistent with working with just one chiral sector on the lattice, as done here.) To lighten the notation, the spinor index ``+'' and the superscript ``max'' will be dropped henceforth.

Arbitrary correlations of fermions in $\A_\Vbb$ can be expressed, via Wick contractions, in terms of the two-point functions $C_{r}$ from \eqref{def C}. The key to calculating the EE is to note that this structure of Wick contractions implies that $\rho_\Vbb$ has the form
\bel{\label{Gaussian rho}
  \rho_\Vbb = \frac1\N \, \e^{- \sum_{x, y \in \Vbb} K_{x, y} \Psi_x\+ \Psi_y}.
}

\newpage

The $M \times M$ matrix $K_{x, y}$ from \eqref{Gaussian rho} can be calculated via \eqref{def rho V} by setting $\O = \Psi\+_x \Psi_y$. Using \eqref{2 point}, this results in the equation
\bel{\label{eq K}
  C_{x, y}
   =
  \frac1\N \Tr\left[
    \Psi_x\+ \Psi_y
    \exp\left\{-\sum_{x', y' \in \Vbb} K_{x', y'} \Psi_{x'}\+ \Psi_{y'}\right\}
  \right].
}
Now assume that $K$ can be diagonalized by a unitary transformation $U$, so that
\bel{
  \left(U\+ K U\right)_{x, y} = \kappa_x \, \delta_{x, y}.
}
Further, define new fermion operators $\~\Psi$ via
\bel{\label{def tilde Psi}
  \Psi_x \equiv \sum_{y \in \Vbb} U_{x, y} \~\Psi_y.
}
Since $U$ is unitary, the fermion commutation relations remain unchanged. Eq.\ \eqref{eq K} can now be expressed as
\bel{
  C_{x, y}
   =
  \frac1\N \sum_{x', y' \in \Vbb}
  U^*_{x, x'}
  \Tr \left[
    \~\Psi_{x'}\+ \~\Psi_{y'}
    \exp\left\{-\sum_{z \in \Vbb} \kappa_z \~\Psi_z\+ \~\Psi_z\right\}
  \right]
  U_{y, y'}.
}
The trace is only nonzero when $x' = y'$, which means that $U^*$ must diagonalize the matrix $C$. Precisely, if $C$ has eigenvalues $\{\varsigma_x\}_{x \in \Vbb}$, they must obey
\bel{
  \varsigma_x = \frac{\e^{-\kappa_x}}{1 + \e^{-\kappa_x}}.
}
(The normalization is fixed by setting $\O = \1$ in \eqref{def rho V}.) In other words, the eigenvalues of $K$ are determined by the eigenvalues of the correlation matrix $C$ via
\bel{\label{def kappa}
  \e^{-\kappa_x} = \frac{\varsigma_x}{1 - \varsigma_x}.
}

This is sufficient to calculate the EE. The von Neumann entropy of a matrix of form \eqref{Gaussian rho} takes the simple form
\bel{
  S_\Vbb = -\Tr[\rho_\Vbb \log \rho_\Vbb]
   =
  -\sum_{x \in \Vbb} \left[
    \varsigma_x \log \varsigma_x + (1 - \varsigma_x) \log (1 - \varsigma_x)
  \right].
}
A simple way to retrieve this result is to note that the $\~\Psi$ modes are all decoupled, so it is enough to sum up the entropies of their individual $2\times 2$ density matrices with eigenvalues $\varsigma_x$ and $1 - \varsigma_x$.

Numerics are needed in the final furlong. Given the explicit form $C_{x, y}$ from \eqref{def C}, its eigenvalues can be numerically calculated for subsystem sizes $M$ spanning several orders of magnitude. Doing this shows that most eigenvalues $\varsigma_x$ are either zero or unity, meaning that they do not contribute to $S_\Vbb$. The number of nontrivial ones scales as $\log M$, and in fact one finds the $M \gg 1$ asymptotics
\bel{\label{S ferm micro}
  S_\Vbb \approx \frac13 \log M.
}
This result is in agreement with continuum techniques, which find the EE of form $S_\Vbb \approx \frac13 \log\frac {L_\Vbb}{a}$, where $L_\Vbb$ is the size of $\Vbb$ in units of a short distance cutoff $a$ \cite{Holzhey:1994we}. This concludes the overview of standard free fermion results.

\newpage

How does this lattice analysis change in the continuum, i.e.\ when one can access only smooth operators? This question can be more precisely restated as follows. Given the operator algebra $\A\^S$ generated by smooth operators $\Psi(x)$ and $\Psi\+(x)$, along with the central elements $\{n_k\}_{k \notin \Pbb\_S}$, what subalgebra $\A_\Vbb\^S$ can be associated to the region $\Vbb$, and what is its ground state entropy?

Since the focus is only on the ground state, it is possible to specialize to the superselection sector in which all particle number operators at $k \notin \Pbb\_S$ satisfy the ``Fermi sea'' condition \eqref{Fermi sea},
\bel{
  n_k = \begin{cases}
          \1, & - \frac N2 \leq k < -k\_S, \\
          0, & k\_S \leq k < \frac N2.
        \end{cases}
}
(Note that the convention to work only with the ``+'' chirality is still in effect.) The only interesting operators from $\A\^S$ that remain are thus the smooth fields $\{\Psi(x), \Psi\+(x)\}_{x \in \Mbb}$, or equivalently the low-momentum modes $\{\Psi_k, \Psi_k\+\}_{k \in \Pbb\_S}$. Either of these two generating sets has $4k\_S$ linearly independent operators.

It is not immediately obvious what subalgebra of $\A\^S$ should be associated to the region $\Vbb$. The issue is clearest if $M$, the number of lattice sites in this region, is large enough so that $M > 2k\_S$. The natural choice in this case is the algebra generated by the $2M$ operators
\bel{\label{Reeh-Schlieder}
  \{\Psi(x), \Psi\+(x)\}_{x \in \Vbb}.
}
However, since $\Vbb$ is larger than $\Pbb\_S$, the above set actually generates the \emph{entire} algebra $\A\^S$! This is the lattice-based version of the famous Reeh-Schlieder theorem \cite{Reeh:1961, Witten:2018zxz}: since smoothness of quantum fields forces them to obey operator constraints, even a region much smaller than $\Mbb$ can support the entire continuum operator algebra. In this case one can reconstruct any momentum mode $\Psi_k$, $k \in \Pbb\_S$ --- or any smooth field $\Psi(y)$, $y \in \Mbb$ --- out of any $2k\_S$ operators $\Psi(x)$, $x \in \Vbb$.

Note, by the by, that the smoothing framework provides a natural limit to the validity of the Reeh-Schlieder theorem. If the subregion is so small that $M < 2k\_S$, the generating set \eqref{Reeh-Schlieder} will \emph{not} have enough linearly independent operators to reconstruct the whole smooth algebra.

This appearance of Reeh-Schlieder implies that the naive algebra one might associate with $\Vbb$ must, in fact, give rise to a reduced density matrix $\rho_\Vbb$ equal to the starting one. Further, since the ground state is an eigenstate of all particle number operators, and they are all preserved under smoothing, the ground state must also remain pure (and invariant) under smoothing. In other words, the ground state entanglement entropy associated to the algebra \eqref{Reeh-Schlieder} must be zero.

It is thus necessary to ``thin down'' the generating set \eqref{Reeh-Schlieder} in order to get a smooth subalgebra that is sensitive to subregion entanglement. One way to do this is to define a special set of coordinates, $\Mbb\_S$, whose elements are
\bel{\label{def MS}
  x_i \equiv \frac{N}{2k\_S} i, \quad 1 \leq i \leq 2k\_S.
}
The scale $\ell\_S \equiv N/2k\_S$ is the ``smearing length'' used to define the continuum fields.

\newpage

Now consider the algebra $\A_\Vbb\^S$ generated by the set
\bel{\label{gen set ferm A}
  \{\Psi(x), \Psi\+(x)\}_{x \in \Mbb\_S \cap \Vbb}.
}
In other words, the generators do not come from every point in $\Vbb$ --- instead, they are spaced a smearing length $\ell\_S$ apart. By the definition of smoothing, fields at points $y \in \Vbb \backslash \Mbb\_S$ can be \emph{approximately} expressed as
\bel{
  \Psi(y) \approx \Psi(x_i) + \frac{y - x_i}{\ell\_S} \big(\Psi(x_{i + 1}) - \Psi(x_i) \big) + \ldots
}
when the point $y$ lies between $x_i$ and $x_{i + 1} = x_i + \ell\_S$. In this sense the algebra $\A_\Vbb\^S$ still probes the field structure at all points in $\Vbb$, but it does not do so to all orders in $k\_S/N$ the way the algebra generated by \eqref{Reeh-Schlieder} did. This algebra has $2M/\ell\_S = \frac MN 4k\_S$ independent generators, so the whole algebra $\A\^S$ can only be reconstructed by taking $\Vbb$ to be approximately the size of the entire space $\Mbb$.

The entanglement entropy associated to this algebra can be found by following the same procedure as in the original lattice calculation. The correlation functions of the smooth fermion fields once again obey Wick's theorem. Indeed, once the superselection sector labels $\{\avg{n_k}\}_{k \notin \Pbb\_S}$ are fixed, for all practical purposes the theory lives on the coarsened lattice $\Mbb\_S$. This means that the desired EE is
\bel{
  S\^S_\Vbb
   \approx
  \frac13 \log\frac M{\ell\_S}
   =
  \frac13 \log \left\{ \frac {2k\_S}{N} M\right\}.
}
As before, after introducing a lattice spacing $a \equiv L_\Vbb/M$, this becomes
\bel{\label{S ferm smooth}
  S\^S_\Vbb \approx \frac13 \log \frac {L_\Vbb}{\ell\_S\^c},
}
where $\ell\_S\^c \equiv a \ell\_S$ is the ``dimensionful'' (or ``continuum'') smearing length.

It is illuminating to compare this result to the microscopic answer \eqref{S ferm micro},
\bel{
  S_\Vbb \approx \frac13 \log M = \frac13 \log\frac{L_\Vbb} a.
}
The only difference lies in the microscopic scale ($a$ vs $\ell\_S\^c$). This simplicity can be misleading. For example, it may seem evident that the replacement $a \mapsto \ell\_S\^c = \frac N{2k\_S} a$ stems from ``integrating out'' high-momentum modes and changing the UV cutoff, as in the standard picture of Kadanoff and Wilson. In fact, smoothing is more subtle than this. Simply removing operators from the algebra was not enough to enact this change: it was also necessary to judiciously pick a new subalgebra $\A\^S_\Vbb$!

To further illustrate this subtlety, recall the oft-cited claim that EE in QFT is divergent because of the singular correlations between fields \cite{Witten:2018zxz}. The above construction strains this narrative. Here the smooth fields still have well defined microscopic correlations, captured by expectations of operators like $\~\Psi\Psi(x, y)$ (cf.\ \eqref{def OPE}) that are $O(1)$ as $x$ approaches $y$ --- i.e.\ that are $O(1/a)$ after the fields are made dimensionful by a rescaling. However, the EE \eqref{S ferm smooth} does not diverge as $\log a$ but rather as $\log \ell\_S\^c$. Thus the divergence of correlators has no direct link with the divergence of the EE. Indeed, $S\^S_\Vbb$ diverges as $M \rar \infty$ without knowing about the rescaling of the $\Psi(x)$'s.

\newpage

The resolution of this tension holds a valuable lesson: there are actually \emph{two} different potential sources of divergence of the EE in a QFT. One comes from genuinely large microscopic correlations between fields. This may happen when the quantities $J_k$ from \eqref{def nk} are all large, e.g.\ in a bosonic theory in which the $\avg{n_k}$'s at low $k$ still scale as, say, $J_k \sim N$. The other source of divergences comes simply from there being a large number of \emph{finite} correlations between fields.

The second kind of singularity is actually the \emph{only} source of EE divergence that is available to a fermionic continuum QFT. Here the $J_k$'s are all manifestly finite and equal to two. Thus for each momentum mode the density matrix must be $2\times 2$ and its eigenvalues must lie between $0$ and $1$. In other words, at each momentum the von Neumann entropy is upper-bounded by $\log 2$, regardless of how one might choose to rescale the fields $\Psi(x)$. It takes a concerted effort of many momenta (or of many fermion species) to obtain a large von Neumann entropy. It is therefore unsurprising that the divergence of EE of the smooth subalgebra $\A_\Vbb\^S$ is set not by the number of lattice sites in $\Vbb$ but instead by the number of independent smooth operators in $\Vbb$, which is the large number $M/\ell\_S = L_\Vbb/\ell\_S\^c$.

This point may seem elementary with hindsight. However, notice that the existence of a second microscopic scale, $\ell\_S$, was crucial to make sense of this explanation. The above distinction between sources of singularities would have been unavailable if one attempted to define a continuum QFT as an effective theory governed by just one microscopic cutoff $a$.

Another technical lesson is that the divergences in the EE within an effective low-energy QFT do not necessarily all scale as $L_\Vbb/a$. In particular, in fermionic theories, these divergences \emph{must} be controlled by the large number $L_\Vbb/\ell\_S\^c$ which is still much smaller than $L_\Vbb/a$.

It is also possible to consider nonmaximal algebras $(\A_\Vbb\^S)^\alpha$ by modifying the generating set \eqref{gen set ferm A}, e.g.\ by removing the generators $\Psi(x_\star)$ and $\Psi\+(x_\star)$ while adding their product $\Psi\+(x_\star) \Psi(x_\star)$ for some $x_\star \in \Mbb\_S \cap \Vbb$. This freedom is directly analogous to the choice between algebras $\A_\Vbb^\alpha$ available in lattice theories \cite{Casini:2013rba, Radicevic:2014kqa, Lin:2018bud}.

Finally, an important moral is that the natural EE of a subregion in a smoothed lattice theory is a low energy phenomenon, in the precise sense that it does \emph{not} depend on the expectations of high-momentum particle numbers $n_k$ or (equivalently) on the ``singular'' behavior that is captured by OPE coefficients. In a conformal theory, the universal coefficient $1/3$ in \eqref{S ferm smooth} is famously equal to $c/3$ where $c$ is the central charge. The above result therefore means that the central charge is completely determined by low-momentum physics. This might sound surprising, but the thought becomes more palatable once you recall that the Kac-Moody current algebra structure in free fermion QFTs is also completely determined by low momentum considerations \cite{Radicevic:2019jfe}. While the Virasoro algebra has not been derived in the smoothing framework yet, there is little doubt that its emergence would closely mimic that of the simpler Kac-Moody algebra in the case of the free Dirac fermion.

\newpage

\section{Free bosons}

The archetypical QFT --- the free scalar field --- presents a much less trivial example of the lattice-continuum correspondence. The large size of the target space means that additional scales must enter the analysis. Consider, for instance, one of the simplest lattice theories that can give rise to the free scalar theory at low energies, the $\Z_K$ clock model at $K \gg 1$. Depending on the exact form of the Hamiltonian, $K$ may compete with the lattice size $N$ and the smoothing scale $k\_S$ in determining the continuum description of this model. In fact, two more scales are necessary in order to get what is normally called the free noncompact (or Gaussian) scalar. These are the target space smoothing scale $p\_S \ll K$ and the target space compactness scale $n\_T \ll p\_S$. The former represents the largest available target momentum, and as such controls the smoothness of wavefunctionals along target space directions. The latter governs the size of small fluctuations witin the subspace of these smooth wavefunctionals. These scales appear as \emph{different} UV cutoffs when the free scalar is viewed as an effective QFT arising from the $\Z_K$ clock model \cite{Radicevic:1D, Radicevic:2D}. Consequently, these scales may all enter the EE of the free scalar.

Here is how this works in $(1+1)$D. The microscopic model consists of $N$ $\Z_K$ clock degrees of freedom arranged on a circle $\Mbb$. The operator algebra on site $x \in \Mbb$ is generated by a \emph{clock operator} $Z_x$ and a \emph{shift operator} $X_x$. They act on basis states $\qvec{\e^{\i\phi_x}}$ as
\bel{
  Z_x \qvec{\e^{\i\phi_x}} = \e^{\i\phi_x} \qvec{\e^{\i\phi_x}},
   \quad
  X_x \qvec{\e^{\i\phi_x}} = \qvec{\e^{\i(\phi_x - \d\phi) }},
}
where $\d\phi \equiv 2\pi/K$. The Hamiltonian is
\bel{\label{def H clock}
  H
   =
  \frac{g^2}{2 (\d\phi)^2} \sum_{x = 1}^N \left(2 - X_x - X_x^{-1} \right) + \frac1{2g^2}\sum_{x = 1}^N \left(2 - Z_x Z_{x + 1}^{-1} - Z_x^{-1} Z_{x + 1} \right).
}

This theory enjoys a Kramers-Wannier duality under the exchange of $X_x$ and $Z_x^{-1} Z_{x + 1}$. This is a strong-weak duality, with the dual coupling being $g^\vee = \d\phi/g$. In the vicinity of the self-dual point $g_\star \equiv \sqrt{\d\phi} \sim 1/\sqrt K$ the low energy states of the theory are governed by a continuum QFT. In fact, a continuum subtheory will exist whenever $1/K \lesssim g \lesssim 1$. This parametric interval is known as the BKT regime \cite{Berezinsky:1970fr, Kosterlitz:1973xp}. The bulk of this paper will focus more specifically on couplings $g \sim 1/\sqrt K$.

A nontrivial fact about this parametric regime is that all low energy states are \emph{tame}. This means that there exists a hierarchy of parameters $K \gg p\_S \gg n\_T \gg 1$ that can be used to define the subspace $\H\^T$ of tame states, and that all states of sufficiently low energy belong to $\H\^T$.

To define $\H\^T$, first consider the $K$-dimensional state space at a single site $x$. (Indices $x$ will be dropped to ease the notation.) This space is spanned by shift eigenstates $\qvec p$ that satisfy $X \qvec p = \e^{\i p\, \d \phi} \qvec p$, with $-K/2 \leq p < K/2$. The states $\{\qvec p\}$ in a narrow band of target momentum, $-p\_S \leq p - p\^{cl} < p\_S$, can be used to define the states $\qvec{\e^{\i \varphi}} \equiv \frac1{\sqrt{2p\_S}} \sum_{p = p\^{cl} - p\_S}^{p\^{cl} + p\_S - 1} \e^{\i p \varphi} \qvec p$. These states have a fixed short distance behavior, in the sense that $\qvec{\e^{\i (\varphi + \d\phi)}} \approx \e^{\i p\^{cl} \d\phi} \qvec{\e^{\i \varphi}}$.

While it is formally possible to define the states $\qvec{\e^{\i\varphi}}$ for all $\varphi \in \R$, only $2p\_S$ of them can be linearly independent. For this reason it will be assumed that $\varphi \in \{0, \d\varphi, \ldots, (2p\_S - 1)\d\varphi\}$ for $\d\varphi \equiv \pi/p\_S$. When $p\^{cl} = 0$, these $2p\_S$ states can be called the \emph{smooth clock eigenstates}, since their wavefunctions $\greek y(\phi) \equiv \qprod{\e^{\i\phi}}{\e^{\i\varphi}}$ necessarily vary slowly as functions of $\phi$. When $p\^{cl} \neq 0$, these states can be called \emph{$p\^{cl}$-modulated clock eigenstates}.

A further reduction of this $2p\_S$-dimensional space comes from restricting to the ``compact'' set of clock eigenstates  $\{\qvec{\e^{\i\varphi}}\}$ for $-\varphi\_T \leq \varphi - \varphi\^{cl} < \varphi\_T$. It is convenient to define the integer cutoff $n\_T \equiv \varphi\_T/\d\varphi \gg 1$. The space spanned by these compact eigenstates is thus $2n\_T$-dimensional. All the wavefunctions $\greek y(\phi)$ belonging to this subspace are compactly supported and smooth (up to a $p\^{cl}$-modulation). This subspace will be called the \emph{tame space at $x$} relative to the \emph{taming backgrounds} $\varphi\^{cl}_x$ and $p\^{cl}_x$. It can be denoted $\H\^T_x(\varphi\^{cl}, p\^{cl})$. The whole tame space $\H\^T$ is now defined as
\bel{
  \H\^T
   \equiv
  \bigoplus_{\{p\^{cl}_x, \varphi\^{cl}_x\}}
  \bigotimes_{x \in \Mbb}
    \H\^T_x(\varphi\^{cl}, p\^{cl}).
}

In practice, the low-lying eigenstates of a model like \eqref{def H clock} will all be tame relative to only a small number of taming backgrounds. The needed taming backgrounds turn out to be solutions to the \emph{classical equations of motion} obtained by separately extremizing the kinetic and potential terms in the Hamiltonian. There is no formal justification of this claim. In this paper it will be simply assumed to be true, and the only taming backgrounds that will be considered will be
\bel{
  p_x\^{cl} = 0, \quad \varphi_x\^{cl} = \varphi\^{cl}\_{const} + \frac{2\pi w}N x,
}
where $w$ is the integer \emph{winding number} and $\varphi\^{cl}\_{const}$ is an integer multiple of $2\varphi\_T$.

States related by $\varphi\^{cl}\_{const} \mapsto \varphi\^{cl}\_{const} + 2\varphi\_T$ are all exactly degenerate. This means that the $\Z_K$ shift symmetry of the model \eqref{def H clock}, generated by the global shift operator $\prod_x X_x$, has a $\Z_{\pi/\varphi\_T} = \Z_{p\_S/n\_T}$ subgroup that is spontaneously broken. This does not contradict the Coleman-Hohenberg-Mermin-Wagner theorem\cite{Mermin:1966fe, Hohenberg:1967zz, Coleman:1973ci} because the coupling $g \sim 1/\sqrt K$ can be safely assumed to be much smaller than the scale $1/\sqrt{\log N}$ below which even $(1 + 1)$D theories can spontaneously break symmetries. This symmetry breaking is somewhat subtle from a purely continuum perspective, as both the shift symmetry and its broken subgroup appear as U(1).

Any operator that preserves $\H\^T$ will also be called \emph{tame}. Projecting to the tame subalgebra will be called \emph{taming}, and the taming of an operator $\O$ will be denoted $\O\_T$. Just like  position-space smoothing, taming does not commute with operator multiplication. All products of tame operators will be understood to be multiplied first and tamed second. The crucial tame operators are the \emph{position} and \emph{momentum} fields
\bel{\label{def phi pi}
  \varphi_x \equiv \frac1{2\i} \left[Z_x \e^{-\i \varphi\^{cl}_x} - Z_x^{-1} \e^{\i \varphi\^{cl}_x} \right]\_T,
   \quad
  \pi_x \equiv \frac1{2\i} \left[X_x - X_x^{-1} \right]\_T.
}
With the above convention for multiplication, these fields satisfy $[\varphi_x, \pi_y] \approx \i \, \delta_{x, y}\1 $ when acting on smooth states in $\H\^T$.

In simple words, the tame subspace contains small fluctuations around fixed classical backgrounds. The tame operators \eqref{def phi pi} are key to constructing the lattice-continuum correspondence for this clock model. Unlike the original fields $Z_x$ and $X_x$, the momentum and position fields have Fourier transforms
\bel{
  \varphi_x \equiv \frac1{\sqrt N} \sum_{k \in \Pbb} \varphi_k \, \e^{\frac{2\pi\i}N kx}, \quad
  \pi_x \equiv \frac1{\sqrt N} \sum_{k \in \Pbb} \pi_k \, \e^{\frac{2\pi\i}N kx},
}
that satisfy the simple relation $[\varphi_k, \pi_l] \approx \i \delta_{k, -l}\1$ when acting on tame states. (Note that in this case the momentum space $\Pbb = \{-\frac N2, \ldots, \frac N2 - 1\}$ \emph{is} the Fourier dual of the original space.) Thus, for $k \neq 0$ and to leading order in the taming parameters, a suitable choice of precontinuum destruction operators is
\bel{\label{def ak}
  a_k \equiv \frac1{\sqrt 2} \left(\frac{\sqrt{\omega_k}}g \varphi_k +  \frac{\i g}{\sqrt{\omega_k}} \pi_k \right), \quad \omega_k \equiv \left| 2\sin \frac{\pi k}N\right|,
}
with $J_k \sim n\_T \gg 1$ for each momentum. The tamed Hamiltonian is
\bel{\label{HT k space}
  H\_T
   \approx
  \frac{g^2}2 \sum_{k \in \Pbb} \pi\+_k \pi_k + \frac1{2g^2} \sum_{k \in \Pbb} \omega_k^2 \varphi\+_k \varphi_k + \frac{2\pi^2}{g^2 N} w^2
   =
  \frac{g^2}2 \pi_0^2
  + \frac{2\pi^2}{g^2 N} w^2 + \!\!\!\!
  \sum_{k \in \Pbb\backslash\{0\}} \!\!\!
    \omega_k n_k
}
for $n_k \equiv a_k\+ a_k$ and up to unimportant additive constants. The ground state is characterized by $\avg{n_k} = 0$ for all $k \in \Pbb \backslash \{0\} = \{-\frac N2, \ldots, -1, 1, \ldots, \frac N2 - 1\}$.

The $k = 0$ mode of the target-momentum field, $\pi_0$, is special. In string theory it is often called just the ``momentum mode,'' while in condensed matter its spectrum is referred to as the ``Anderson tower of states'' \cite{Anderson:1952}. It is a tame degree of freedom that can be associated to a kind of coherent collective propagation of all clocks on the chain $\Mbb$ --- sometimes a bit sloppily described as the ``center of mass'' momentum in the target space. It is also a constant of motion, as $\pi_0$ commutes with all other operators to first order in the taming parameters, and the ground state is its null state. A somewhat nontrivial fact, implied by the self-consistency of the tameness assumption, is that the eigenvalues of $\pi_0$ are integer multiples of $K/\sqrt N$ \cite{Radicevic:2D}.\footnote{
  It is worth stressing that Kramers-Wannier duality (or T-duality in the string literature) exchanges these integers with the winding numbers $w$. Further, the largest possible momentum or winding number is set by the taming and smoothing parameters. For example, if $\varphi\^{cl}_x$ is required to vary slowly, such that it does not change by more than $2\varphi\_T$ over the smoothing scale $\ell\_S = N/2k\_S$, the winding number becomes constrained to obey $|w| \lesssim \varphi\_T k\_S$.
}

The smoothing procedure can now be applied by projecting away all ladder operators at momentum $|k| \notin \Pbb\_S = \{-k\_S, \ldots, k\_S\}$. This leads to smooth fields
\bel{
  \varphi(x) \equiv \frac1{\sqrt N} \sum_{k \in \Pbb \backslash \{0\}}
    \varphi_k \, \e^{\frac{2\pi\i}N kx}, \quad
  \pi(x) \equiv \frac1{\sqrt N} \sum_{k \in \Pbb  \backslash \{0\}}
    \pi_k \, \e^{\frac{2\pi\i}N kx}.
}
The $k = 0$ modes are treated separately because, as just discussed, the operator $\pi_0$ is not used to create precontinuum operators $a_k$ from \eqref{def ak}. This can be understood as the constraint $\sum_{x \in \Mbb} \varphi(x) = \sum_{x \in \Mbb} \pi(x) = 0$ imposed on smooth fields.

The natural smooth subalgebra associated to a region $\Vbb$ is generated by
\bel{\label{gen set bos A}
  \{\varphi(x), \pi(x)\}_{x \in \Mbb\_S \cap \Vbb},
}
where $\Mbb\_S$ is the coarse lattice consisting only of $2k\_S$ points $x_i$ \eqref{def MS}. This is directly analogous to the smooth subalgebra \eqref{gen set ferm A} in a fermion theory, with one caveat. The subalgebra $\A\^S_\Vbb$ generated by the set \eqref{gen set bos A} is completely independent of the $\pi_0$ mode, which does not figure in the definition of any of the operators $\pi(x)$.

Now consider the density matrix associated to this subalgebra. Just like in the fermionic case, the idea is to exploit the fact that correlation functions satisfy Wick's theorem. Although this theorem cannot apply to bosonic $n$-point correlators when $n$ is comparable to any of the taming or smoothing parameters, the fact that it holds for few-point correlators means that a natural Ansatz for the reduced density matrix in the continuum theory, to leading order in taming/smoothing parameters and restricted to acting only on tame states, is
\bel{
  \rho\^S_\Vbb
   \propto
  \e^{- \sum_{x, y \in \Mbb\_S \cap \Vbb} \left[
    K_{x, y}\^S \varphi(x) \varphi(y) + \~K_{x, y}\^S \pi(x) \pi(y) \right] },
}
with the matrices $K\^S$ and $\~K\^S$ chosen so that, in momentum space,
\bel{\label{ansatz rho V S}
  \rho\^S_\Vbb
   =
  \frac1{\N\^S} \, \e^{- \sum_{k, l \in \Pbb\_S \backslash \{0\}}
    A_{k, l}\^S a_k\+ a_l }.
}
The analogous Ansatz on the lattice (with the subregion algebra generated by tame but non\-smooth fields $\varphi_x$ and $\pi_x$) would have been
\bel{
  \rho_\Vbb
   \propto
  \e^{- \sum_{x, y \in \Vbb} \left[
    K_{x, y} \varphi_x \varphi_y + \~K_{x, y} \pi_x \pi_y
  \right] },
}
with an additional relation between $K$ and $\~K$ that guarantees the simple form
\bel{\label{ansatz rho V}
  \rho_\Vbb
   =
  \frac1{\N} \, \e^{- \sum_{k, l \in \Pbb \backslash \{0\}}
    A_{k, l} a_k\+ a_l }.
}

The $A$-matrices in the Ans\"atze \eqref{ansatz rho V S} and \eqref{ansatz rho V} depend on the size of $\Vbb$ and can be expressed in terms of the correlation matrices
\algnl{
  F_{x, y}
   &\equiv
  \avg{\varphi_x\varphi_y}
   =
  \frac{g^2}{N} \sum_{k \in \Pbb \backslash \{0\}}
    \frac1{2\omega_k} \e^{\frac{2\pi\i}N k(x - y)}, \\
   \quad
  P_{x, y}
   &\equiv
  \avg{\pi_x \pi_y}
   =
  \frac{1}{g^2 N} \sum_{k \in \Pbb \backslash \{0\}}
   \frac{\omega_k}{2} \e^{\frac{2\pi\i}N k(x - y)},
}
or their smooth (i.e.\ ``normal-ordered'') versions
\algnl{
  F_{x, y}\^S
   &\equiv
  \avg{\varphi(x)\varphi(y)}
   =
  \frac{g^2}{N} \sum_{k \in \Pbb\_S \backslash \{0\}}
    \frac1{2\omega_k} \e^{\frac{2\pi\i}N k(x - y)}, \\
   \quad
  P_{x, y}\^S
   &\equiv
  \avg{\pi(x) \pi(y)}
   =
  \frac{1}{g^2 N} \sum_{k \in \Pbb\_S \backslash \{0\}}
   \frac{\omega_k}{2} \e^{\frac{2\pi\i}N k(x - y)}.
}
This computation (and the subsequent evaluation of the von Neumann entropy) parallels the fermionic one, but is much more involved. In particular, the eigenvalues of $A$ or $A\^S$ cannot be \emph{directly} determined from the eigenvalues of $F$ and $P$ via equations like $F_{x, y} = \Tr[\rho_\Vbb \varphi_x \varphi_y]$.

\newpage

Here is an elegant way to proceed instead \cite{Casini:2009sr}. Focus on the lattice calculation, for simplicity; the smoothed calculation requires you to substitute $\Vbb$ by $\Vbb \cap \Mbb\_S$ in what follows. Let $\~a_z$, $z \in \Vbb$, be the bosonic ladder operators such that
\bel{
  \sum_{k, l \in \Pbb \backslash \{0\}}
    A_{k, l} a_k\+ a_l
   =
  \sum_{z \in \Vbb}
    \alpha_z \~a_z\+ \~a_z.
}
This is completely analogous to the fermionic operators $\~\Psi_z$ introduced in \eqref{def tilde Psi}. Define the square, not necessarily unitary matrices $U$ and $V$ such that
\bel{
  \varphi_x \equiv \sum_{y \in \Vbb} \left(
   U_{x, y} \~a_y +  U_{x, y}^* \~a_y\+
  \right),
   \quad
  \pi_x \equiv \sum_{y \in \Vbb} \left(
    V_{x, y} \~a_y +  V_{x, y}^* \~a_y\+
  \right).
}
The $F$ and $P$ matrices can now be expressed as
\algnl{\notag
  F_{x, y}
   &=
  \Tr[\rho_\Vbb \varphi_x \varphi_y] \\ \notag
   &=
  \frac1\N
  \sum_{z', z'' \in \Vbb}
    \Tr\left[
      \e^{-\sum_{z \in \Vbb} \alpha_z \~a_z\+ \~a_z}
      \left(
       U_{x, z'} U^*_{y, z''} \~a_{z'} \~a_{z''}\+
        +
       U_{x, z'}^* U_{y, z''} \~a_{z'}\+ \~a_{z''}
      \right)
    \right] \\ \label{def F}
   &\equiv
  \left(U^*\, \~n\, U\^T \right)_{x, y}
   +
  \left(U \, (\1 + \~n)\^T\, U\+\right)_{x, y}
}
and
\bel{\label{def P}
  P_{x, y}
   =
  \left(V^*\, \~n\, V\^T \right)_{x, y}
   +
  \left(V \, (\1 + \~n)\^T\, V\+\right)_{x, y},
}
where
\bel{
  \~n_{x, y}
   \equiv
  \Tr \left[\rho_\Vbb \~a_x\+ \~a_y \right]
   \approx
  \frac{\delta_{x, y}}{\e^{\alpha_x} - 1}.
}
This last result holds to leading order in $J_k$, and hence to leading order in $n\_T$.

The canonical commutation relation implies that $UV\^T = -\frac12 \1$, and hence multiplying eqs.\ \eqref{def F} and \eqref{def P} gives
\bel{
  (FP)_{x, y} = \left[ U \left(\~n + \tfrac12 \right)^2 U^{-1} \right]_{x, y}.
}
In other words, the eigenvalues of $FP$ --- and not of $F$ or $P$ --- are related in a simple way to the eigenvalues of $A$. Specifically, if the eigenvalues of $\sqrt{FP}$ are denoted $\{\varsigma_x\}$, the spectrum of $A$ is approximately given by
\bel{
  \e^{-\alpha_x}
   \approx
  \frac{\varsigma_x - 1/2}{\varsigma_x + 1/2}.
}
Compare this to the fermionic ``entanglement spectrum'' $\{\kappa_x\}$  given by eq.\ \eqref{def kappa}. The entropy is
\bel{
  S_\Vbb
   =
  \sum_{x \in \Vbb} \left[
   (\varsigma_x + \tfrac12) \log(\varsigma_x + \tfrac12)
   - (\varsigma_x - \tfrac12) \log(\varsigma_x - \tfrac12)
  \right].
}

The spectrum $\{\varsigma_x\}_{x \in \Vbb}$ is found by diagonalizing the matrix
\bel{
  (FP)_{x, y}
   =
  \frac1{4N}
  \sum_{k, l \in \Pbb\backslash\{0\}}
    \frac{\omega_l}{\omega_k} \e^{\frac{2\pi\i}N(kx - ly)} d(k - l),
  \quad
  d(k - l)
   \equiv
  \frac1N \sum_{z \in \Vbb} \e^{\frac{2\pi\i}N (l - k) z}.
}
In the special case $\Vbb = \Mbb$, $d(k - l)$ becomes a $\delta$-function, giving $FP = \frac14 \1$ and immediately implying $\sqrt{\varsigma_x} = 1/2$ for all $x \in \Vbb$, resulting in $S_\Vbb = 0$.
\newpage

When the region $\Vbb$ is much smaller than the entire space $\Mbb$ --- the standard choice! --- one must again resort to numerics. The calculation is subtle, but there are many ways to approach it, and it is rather well understood in the literature \cite{Plenio:2004he, Casini:2007bt, Casini:2009sr, Metlitski:2011pr}. When $\Vbb$ contains $M$ sites, with $1 \ll M \ll N$, the main points are:
\begin{romanlist}
  \item Keeping only the dominant $M$-dependent term, the EE is
    \bel{
      S_\Vbb \approx \frac13 \log M.
    }
    This is consistent with the fact that this is a conformal theory with total central charge unity. In continuum notation, this result is
    \bel{
      S_\Vbb \approx \frac13 \log \frac{L_\Vbb}{a}.
    }
  \item A more precise analysis reveals a subleading term that also depends on $M$. With it, the EE appears to have the form
      \bel{\label{double log S}
        S_\Vbb \approx \frac13 \log M + \frac12 \log\log \frac N {2\pi M}.
      }
      In the literature one typically encounters this term by performing purely continuum manipulations and introducing an ad hoc infrared regulator $m$, a mass scale that governs the smallest energy gap in the system; at the end of the day, a term of the form $\frac12 \log(-\log(m L_\Vbb))$ appears in the EE \cite{Casini:2009sr}. On the lattice there actually exists a nonzero gap, and it is equal to $2\pi/N$. It is therefore natural to conjecture that the above formula is correct. Unfortunately, it is difficult to verify this conjecture by numerically extracting the double logarithm.
  \item The coefficient $1/2$ in eq.\ \eqref{double log S} is a lesser known (but nevertheless universal) cousin of the central charge. It is important in higher dimensions, in particular in $(2 + 1)$D. There the EE of a free scalar appears to have the form\cite{Metlitski:2011pr, Agon:2013iva}
      \bel{
        S_\Vbb\^{(3D)} =  c_1 M + c_0 + \frac{c_{-1}}M + \ldots + \frac12 \log\left\{g^2 M \log \frac N{2\pi M}\right\}.
      }
      The powers of $M$ (or, in continuum notation, of $L_\Vbb/a$) are well known, and the quantity $c_0$ (often called $F$) is universal. The last term in this expression is needed to ensure consistency with particle-vortex duality. Namely, the EE of a weakly coupled gauge theory dual to this scalar must contain a term of form $\frac12 \log\{g^2 M\}$ that can smoothly cross over to the topological EE $\frac12 \log \kappa$ of a U(1)$_\kappa$ Chern-Simons theory, as such a theory may always be present in the infrared, at lengths greater than $\kappa/g^2$ \cite{Agon:2013iva, Radicevic:2015sza}.\footnote{
        The ``infrared'' nature of this logarithmic term may at first glance appear to be connected with the other familiar infrared degree of freedom, the momentum mode $\pi_0$. There is no direct connection, however. The mode $\pi_0$ is simply absent from the subregion algebras $\A_\Vbb$ and $\A_\Vbb\^S$. It cannot contribute to the EE. However, this additional logarithmic entropy can be understood to come from the ``zero-momentum'' mode within the region $\Vbb$ itself \cite{Radicevic:2015sza}. In simplistic terms, every mode of wavelength greater than $M$ will register as a constant mode in the region $\Vbb$, and hence this mode will have an enhanced entropy.
      }
\end{romanlist}

An entirely analogous calculation can be carried out using just the smooth subregion algebra $\A_\Vbb\^S$. All the results remain qualitatively the same; as in the fermionic case, the only difference is that the momentum space cutoff is now $M/\ell\_S = L_\Vbb/\ell\^c\_S$ instead of $M = L_\Vbb/a$. In particular, the smooth subregion algebra remains sensitive to the same universal coefficients as the lattice subregion algebra.

The upshot of this analysis is that the EE of a scalar field has two sources of divergence as the number of lattice sites is taken to infinity. The $\log M$ divergence is, just like in the fermionic case, simply due to a large number of modes that each have a finite EE. The subleading divergence, exemplified by the $\log(-\log M)$ term in $(1 + 1)$D, instead appears to come from a single mode that corresponds to the collective (``center of mass'') motion of all the clocks within $\Vbb$. Importantly, the fact that correlation functions of continuum fields $\varphi(x)$ behave as $\log|x - y|$ has no connection to the dominant divergences of the EE.

\section{Conclusion}

This was a brief, algebra-based review of entanglement entropy in free quantum field theories. While the calculations hewed closely to the existing literature, a few novel points were still made. The principal novelty was the construction of the smooth subregion algebras $\A_\Vbb\^S$ that involve only continuum quantum fields, avoid the pitfall of the Reeh-Schlieder theorem, and reproduce the well known divergence structure of the EE. Another new point was that the dominant divergences in the EE have no connection to divergences in continuum correlation functions, and instead owe their existence to a large number of modes that each contribute a \emph{finite} entropy. These new points notwithstanding, this paper provides a streamlined calculation of EE that may be of use as a reference on its own.

A large number of generalizations of the calculations presented here can now be pursued. Free theories in higher dimensions and with the addition of gauge fields may be studied using the same methods. Interactions can be included perturbatively. Entropies of nonmaximal algebras associated to a region $\Vbb$ can be studied as well. Other entanglement measures, such as mutual informations, Renyi entropies, min-entropies, etc.\ are also worth studying. Finally, all these results can be further refined by calculating corrections due to the finiteness of various smoothing and taming parameters.

\section*{Acknowledgments}

It is a pleasure to thank Tom Faulkner and Matt Headrick for useful discussions, and Ruben Verresen, Ryan Thorngren, and John McGreevy for asking the right questions about the Reeh-Schlieder theorem. This work was completed with the support from the Simons Foundation through \emph{It from Qubit: Simons Collaboration on Quantum Fields, Gravity, and Information}, and from the Department of Energy Office of High-Energy Physics grant DE-SC0009987 and QuantISED grant DE-SC0020194.

\bibliographystyle{ws-mplb}
\bibliography{Refs}

\end{document}